# Multifractality in sputtered crystalline and amorphous transition metal nitride thin films

**K. Anantha Padmanabhan[1], Kuraganti Vasu[2], P. Manimaran[3], M. Ghanashyam Krishna[4,*]**

[1] Materials Science and Engineering Division, Department of Mechanical Engineering, College of Engineering, Anna University, Guindy, Chennai, 600025, India

[2]Department of Physics, School of Advanced Sciences, Vellore Institute of Technology, -Vellore-632014, Tamil Nadu, India

[3]School of Physics, [4]Centre for Advanced Studies in Electronics Science and Technology, University of Hyderabad, Hyderabad, Telangana, 500 046, India

*Corresponding author (Email: mgksp@uohyd.ac.in )

## Abstract

In this work, we present evidence for multifractal growth in Titanium and Niobium nitride thin films on different substrates. Self-similar multifractal structures such as dendrites and multi-branched snowflake-like structures are reported. The fractal dimensions are controlled by a combination of processing parameters and conditions at the film-substrate interface. Significantly, fractal growth is observed not only in crystalline films on crystalline and amorphous substrates but also on amorphous films on crystalline substrates. The 2D Multifractal Detrended Fluctuation Analysis (2D MFDFA) carried out on the scanning electron micrographs, clearly reveals the multifractal nature of the nitride thin films. The statistical moments were varied between -5 to +5. The Hausdorff dimensions for the TiN films are in the range of 1.65 to 1.79 while it is 1.64 to 1.74 for the NbN thin films. It is proposed that the observed multifractality is due to a competition between morphological instabilities and Diffusion-Limited Aggregation.

## 1. Introduction

Fractal structures of both crystalline and amorphous nature have created much interest in recent years due to their implications for applications and the insights they provide into the fundamental aspects of thin film (2D) and crystalline and amorphous solid (3D) growth [1-5]. The earlier literature on fractals in thin films was mainly focused on fractals-based analysis of surface texture and roughness in the context of specific applications. [5-16], while the more limited research is aimed at analyzing the fractal-like dendritic structures observed in thin films [17-22]. The dendritic structures are very interesting since they lead to an understanding of the fundamental concepts of crystallization in thin films on a variety of substrates [21,22] and they are also found in 3D crystal growth.

Many natural and man-made phenomena ranging from climate changes, geophysical processes, formation of snowflakes to dendrites in materials, stock markets, sociological aspects of life and heritage architecture can be described within the framework of Non-linear dynamics or chaos dynamics. [1-5].

A survey of literature reveals that multifractality is present in a variety of elemental, metal oxide and metal nitride thin films. Two types of multifractal objects are considered: one based on surface texture and the other on the formation of dendritic structures, as in the present case. For example, Jurecka et al. [12] have reported multifractality in Si layers. Shakoury et al. [15] have reported this in ZnS films deposited by RF magnetron sputtering with the width of the singularity spectrum being dependent on processing conditions. Astinchap et al. have demonstrated multifractality in MOCVD deposited $TiO_2$ thin films [16]. It is pertinent to note that the fractal analysis is carried out on AFM images displaying surface roughness of the films in these cases. Hui et al. [17] have grown snowflake-like branched multifractal structures in $Ag_2O$ thin films. More significantly, the growth of these structures is attributed to diffusion-limited aggregation. Schwarzer et al. [18] have investigated the multifractal properties of some diffusion-limited aggregates. Sanchez et al. [19] have reported geometrical multifractal patterns in laser irradiated Ge/Al multilayer thin films. There is work on metal oxide thin films from which the possibility of fractality can be inferred [20,21,22]. In the specific case of binary transition metal nitrides such as TiN, it is well documented that the microstructural evolution is strongly dependent on process parameters such as temperature and pressure [23]. Kinetic Monte Carlo and molecular dynamics studies have been used to predict the observed microstructural evolution as a function of process parameters such as growth rate, temperature etc. [24]. In addition, fractal behaviour is also reported in TiN thin films[25-27]. However, in contrast to the present study, these do not report dendritic or snowflake like structures The current authors have demonstrated that crystalline dendritic structures can also grow on amorphous substrates by the process of diffusion limited aggregation induced by the elastic strain at the film-substrate interface [28, 29]. Titanium and Niobium nitride (TiN and NbN) thin films were employed as the prototypical material systems to illustrate this point.

In the present work a detailed fractal geometry-based analysis of the dendritic and snowflake structures observed in sputter deposited TiN and NbN thin films is presented. It is shown that the multifractal structures can be achieved over a diverse range of parameters, on different substrates (glass and single crystal Si), varying substrate temperatures, thicknesses and nature of crystallinity in the films (i.e. amorphous or crystalline).

## 2. Experimental

The details of thin-film growth are presented in our previous works [28,29]. TiN and NbN thin films were deposited on glass and Si substrates by RF magnetron sputter deposition at a power of 100 W. The sputtering targets were Ti and Nb of 2-inch diameter respectively and the gas atmosphere was 100% Nitrogen at a pressure of $5x10^{-2}$ mbar while the base pressure was $3 \times 10^{-6}$ mbar. The substrates (glass and silicon) were heated between 250 and $300^{0}$C during the deposition process while the interelectrode distance was kept as 4-6 cm. In addition to chemical cleaning [28], the Si substrates were subjected to a plasma cleaning step, during pre-sputtering, to avoid surface oxidation. Thin films of different thicknesses ( 40-300 nm) were grown by varying the deposition time between 60 and 240 minutes. As a control experiment, Ti thin films were deposited on single crystal Si substrates by RF magnetron sputtering at room temperature and 100 W power. The gas atmosphere was 100% pure Ar while the target-substance distance was 6 cm and the total deposition time was 60 minutes. The thickness of the films was determined by a surface profilometer (Model XP-1, Ambios Technology, USA). The crystal structure of TiN films was determined using x-ray diffractometer equipped with Co-$K_\alpha$ ($\lambda$= 1.7889A$^o$) radiation. Surface morphology of the films was determined by field emission scanning electron microscopy (Carl Zeiss, ultra 55) and contact mode atomic force microscopy (Seiko Inst, SPA-400, Japan).

In the present work 2D Multifractal Detrended Fluctuation Analysis (MFDFA) was employed for quantitave extraction of fractal parameters. To ensure reliability, all images were pre-processed prior to 2D MFDFA analysis, including contrast enhancement, noise reduction, and segmentation to isolate the region of interest. Additionally, prior to applying 2D Multifractal Detrended Fluctuation Analysis (MFDFA), the SEM images were segmented to extract the region of interest (ROI) and eliminate background or artifacts. This segmentation step ensures that only the relevant surface features contribute to the multifractal spectrum. These steps minimize artifacts and allow meaningful and reproducible extraction of multifractal parameters ($\Delta\alpha$, $H$, $D_f$ :to be defined in the next section).

MATLAB's built-in function "imbinarize" was used to convert SEM images into binary form for fractal analysis using the box-counting method.

Table 1: Deposition conditions of the TiN and NbN thin films and corresponding sample names

| S. No. | Sample name | Material | Substrate | Substrate temperature | Thickness | Image details |
|---|---|---|---|---|---|---|
| 1 | 1, 2 and 3 | TiN | Glass | 300°C | 150 nm | Different locations and magnifications on the same sample |
| 2 | 4 | TiN | Single crystal Si | 250°C | 250 nm | |
| 3 | Sample 5 and 6 | NbN | Single crystal Si | 250°C | 300 nm | Different locations and magnifications on the same sample |

## 3. Theory and Calculations

The theoretical framework for fractal behaviour has been described in detail earlier [1-5] and only the relevant background is provided here.

The monofractal dimension also known as the Hausdorff index or dimension, $D_f$, is calculated using the box counting method [30-33]. If the number of boxes of size '$a$' required to cover a fractal object is $N(a)$ then the fractal dimension, $D_f$ is defined by eqn (1):

$$D_f = \frac{\ln N(a)}{-\ln a} \qquad (1)$$

Eqn (1) implies that

$$N(a) \propto a^{-D_f} \qquad (2)$$

Significantly, it is not necessary for $D_f$ to be an integer number. However, for real fractal objects the boxes need not be of uniform size and therefore, for boxes of varying sizes eqn (1) is re-written as

$$D_f = \lim_{a \to 0} \frac{\ln N(a)}{-\ln(\frac{1}{a})} \qquad (3)$$

For the multifractal analysis, the probability density of the $i^{th}$ box is considered as

$$P_i(l) = \frac{N_i(l)}{\sum N_i(l)} \qquad (4)$$

where $N_i(l)$ is the number of pixels containing mass in the $i^{th}$ box of size $l$ $(l \leq 1)$ and the denominator indicates the total mass of the considered system in Eq. (4). Holder exponent $\alpha_i$ and probability density $P_i(l)$ are related as

$$P_i(l) \propto l^{\alpha} \qquad (5)$$

The Holder exponent in Eq. (5) becomes monofractal when the nonlinearity in mass distribution between the squares is absent. A distribution of Holder exponent values could be found in a system. The number of squares N(α) where the probability density $P_i$ has $\alpha_i$ values between $a$ and $a+da$ can be defined as

$$N(\alpha) \propto e^{-f(\alpha)} \qquad (6)$$

where $f(a)$in Eq. (6) is called multifractality spectrum. Multifractality in terms of $q^{th}$ order probability density distribution terms is

$$z(q,l) = \sum_{i=1}^{N(l)} P_i^q(l) \sim l^{(q-1)D_q} \qquad (7)$$

In Eq. (7), $D_q$ is the generalized fractal dimension and $(q-1)D_q$ is called the mass exponent and is represented as $\tau(q)$. It is obtained from the slope of the log-log plot between $z(q, l)$ vs $l$. The mass exponent and the multifractality are related by the Legendre transformation as

$$F(\alpha, q) = q\alpha(q) - \tau(q) \qquad (8)$$

also

$$\alpha(q) = \frac{d\tau(q)}{dq} \qquad (9)$$

in Eq. (8). The graph between $f(\alpha)$ vs $\alpha$ in the multifractal spectrum shows the strength of the multifractality of the data under study.

The Wavelet Transform Modulus Maxima (WTMM), Sandbox, lacunarity and 2D Multifractal Detrended Fluctuation Analysis (2D MFDFA) methods are frequently used for analysis of fractals and multifractals [30-32]. The WTMM method is well suited for continuous data and edge detection but tends to be computationally expensive and highly dependent on the choice of wavelet. The Sandbox method, while simple to apply, is prone to statistical noise and edge effects, which can limit its reliability. Lacunarity analysis is effective for capturing texture and heterogeneity within a dataset, yet it does not yield the complete multifractal spectrum. In contrast, the 2D Multifractal Detrended Fluctuation Analysis (2D MFDFA) method offers a more balanced approach, as it is robust, resistant to trends, numerically stable, capable of providing the full multifractal spectrum, and performs well even with finite or noisy data [33-36]. It is an advanced version of the 1D MFDFA method and helps identify whether patterns repeat across different scales (scale-invariance) and whether they have complex, multi-level structures (multifractality). This technique is commonly used in areas such as image texture analysis, surface roughness in materials, terrain modelling, and biological imaging. Hence, it has been used in the present study.

In this analysis an image under study can be considered as a self-similar surface having domain size $(M \times N)$ which can be denoted using a two-dimensional array

$$X(i,j)\,, i = 1,2, \dots, M \text{ and } j = 1,2, \dots, N. \qquad (10)$$

The cumulative sum can be written as

$$G(i,j) = \sum_{k_1=1}^{i} \sum_{k_2=1}^{j} X(k_1\,, k_2) \qquad (11)$$

Then $G(i, j)$ can be divided into $M_s \times N_s$ such that $M_s \equiv \frac{M}{s}$ and $N_s \equiv \frac{N}{s}$.

Where square subdomains which do not overlap and have length, $s$, each. The detailed procedure of the 2D-MFDFA method employed in this study is described in [26].

If $F_q(s)$ is the $q^{th}$ order fluctuation function, then the graph between $ln(F_q(s))$ $vs$ $ln(s)$ for each value of $q$ can be analyzed and the scaling behaviour can be determined. The power-law correlation for $G_{m,n}$ for long-range can indicate the increase in $F_q(s)$for higher values of $s$.

$$F_q(s) \propto s^{h(q)} \tag{12}$$

The relation between the Hurst index (h(2)) with fractal dimension $D_f$ is $h = 3 - D_f$ [37, 38]

The scaling function $F_q(s)$ is calculated for different values of $q$ and plotted on a log-log graph. If the plot shows a straight line over a range of scales, it means that the data has reliable scaling behaviour. The slope of this line gives the generalized Hurst exponent $h(q)$, which shows how different types of fluctuations behave at different scales. When $h(q)$ is independent for all values of $q$, the data is called monofractal, meaning it has a uniform structure. If $h(q)$ changes with $q$, the data is multifractal, showing a more complex structure. A decreasing $h(q)$ with increasing $q$ suggests that large fluctuations are less correlated than smaller ones. The Hurst exponent $H$, often defined as $h(2)$, helps describe correlation behaviour in the data and values of:

- $H < 0.5$ show anti-persistent behaviour,
- $H = 0.5$ means uncorrelated or random behaviour, and
- $H > 0.5$ indicates persistent or positively correlated behaviour.

Another way to understand multifractality is through the multifractal spectrum, a plot of $f(\alpha)$ against $\alpha$, where α represents the strength of singularities (sharp changes), and $f(\alpha)$ is the fractal dimension of points with those singularities. The width of this spectrum (difference between maximum and minimum $\alpha$ shows how multifractal the data is, i.e. a wider spectrum means stronger multifractality, while a narrow one indicates weaker multifractality.

In MFDFA method, $q < 0$ emphasizes small fluctuations and $q > 0$ emphasizes large fluctuations, where $q$ takes integer values. In our study, we considered that the range [-5, +5] is wide enough to capture multifractal behaviour from both rare small fluctuations and dominant large fluctuations, which is the main goal of MFDFA. Increasing $q$ beyond ±5 does not significantly add new insights, but it increases computational error and destabilizes the scaling exponent $h(q)$.

The regions of interest (ROIs) were identified by examining the SEM images of the films deposited under different conditions given in Table 1. The different locations on the images were selected and analysed to avoid artefacts and edge effects. The observed multifractal behaviour and scaling characteristics were consistent across these regions. Most importantly, the current dataset was primarily optimized for nanoscale characterization rather than large-area imaging; therefore, such mosaics were not systematically acquired.

## 4. Results and Discussion

All the TiN films are crystalline independent of the substrate on which they are deposited whereas the NbN is amorphous [28, 29]. It is significant that the TiN films deposited on amorphous glass substrates are also crystalline.

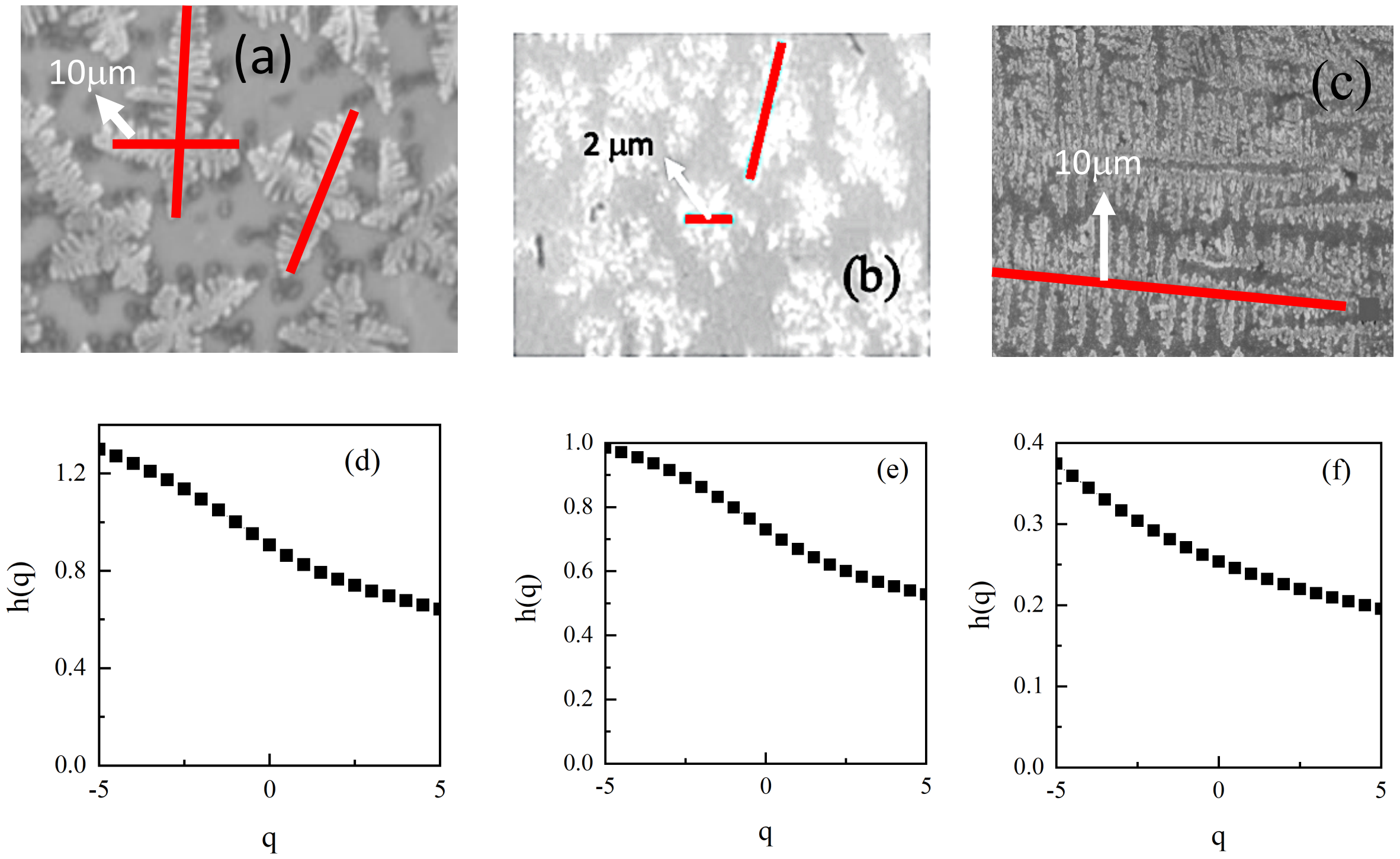

Figure 1: FE-SEM images of the TiN films of 150 nm thickness deposited on glass substrate at a substrate temperature of 300°C, (a) sample 1 at a magnification of 3.22 KX, (b) sample 2 at a magnification of 11.44 KX and (c) sample 3 at a magnification of 11.59 KX. The variation in the corresponding Hurst exponents h(q) as a function of the moments, q is shown in (d), (e) and (f) respectively.

The FE-SEM images of the 150 nm thickness TiN film deposited on a glass substrate at 300°C designated as samples 1,2 and 3 respectively are shown in fig 1(a), (b) and (c) at different locations and magnifications. The samples displayed a classical fractal structure which is a combination of dendrites and snowflakes. The dendrites have at least three scales of self-similar structures. At the macroscale the main branch length was of the order of 50 μm with the widest secondary branch 10 μm wide. Several other smaller secondary branches of sub-micron and nanometric width were also visible. The FE-SEM image of the 250 nm thickness TiN films deposited on single crystal Si substrates at a substrate temperature of 250°C, in Fig 2(a), displayed a snowflake-like geometry. The multi-branched object has the longest dimension of the order of 5 μm and several smaller self-similar branches which were sub-micron to nanometric dimensions. As described in the experimental section a thin film of metallic Ti was also deposited under similar conditions, to ensure that these morphologies are a consequence of nitridation of the metal. The SEM image displayed in Fig. 2(c) clearly indicates the complete absence of fractal structures. It is, therefore, inferred that the fractal structures observed in TiN films are a consequence of nitridation. It is pointed out that the low magnification images of samples 4,5 and 6 were presented in ref [28] and hence they are not reproduced here. The self-similarity of these structures at all the scales is evident from the SEM images. Interestingly, the amorphous NbN film deposited on a single crystal Si substrate also displays a multi-branched snowflake structure with several sub-branches that are of nanometre dimensions as seen from Figs. 3(a) and (b). The more detailed descriptions of these structures are presented in our previous works [28, 29].

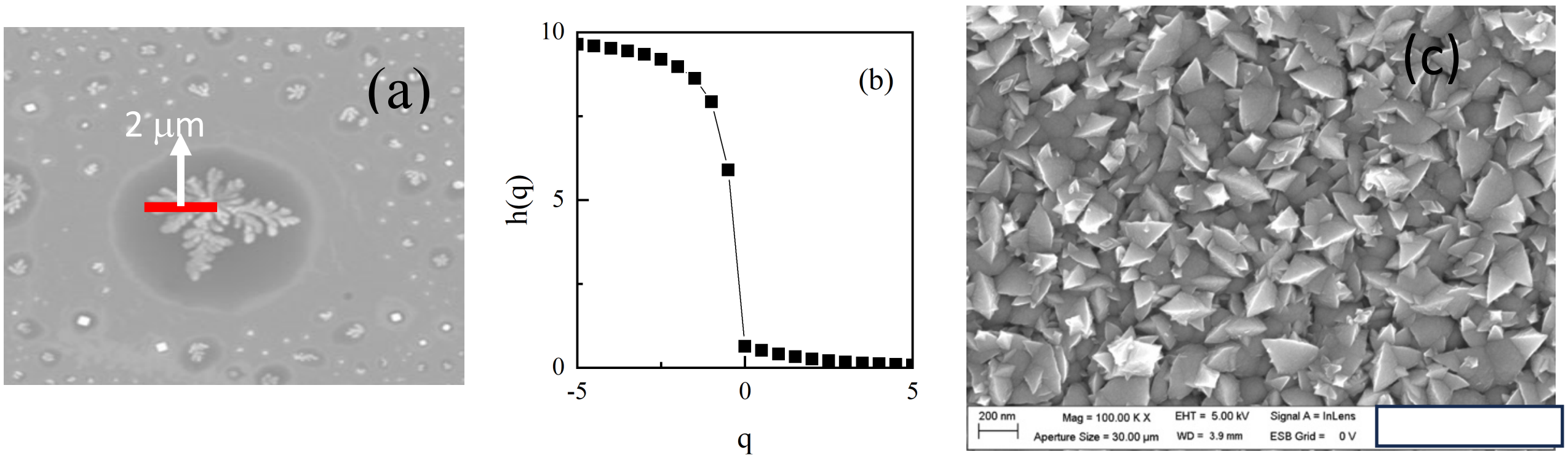

Figure 2: (a) FE-SEM image of the TiN films of 250 nm thickness deposited on a single crystal Si substrate at a substrate temperature of 250$^{o}$C, designated as sample 4 at a magnification of 20.09 KX. The variation in the corresponding Hurst exponents h(q) as a function of the moments ,q is shown in (b) and (c) is the SEM image of a Ti thin film.

The Hurst exponents *h(q)* in figs 1 to 3 estimated from the equations given in the previous sections are plotted as a function of the statistical moment *q*, in the range -5 to +5. They are displayed in Figs. 1(d), (e) and (f) for samples 1, 2 and 3 respectively. The corresponding plots for samples 4, 5 and 6 are shown in figs 2(b), 3(c) and (d) respectively and the values are presented in Table 2. The Hausdorff dimensions are also presented in table 2. The mass exponents $\tau(q)$ *vs* $q$ plots for the same samples are displayed in Figs. 4(a) and(b) and fig. 5 respectively, The Hurst exponents indicate that samples 1, 2 and 5 exhibit persistent long-range correlation, sample 4 shows random or no correlation whereas samples 3 and 6 exhibit anti-persistent behaviour.

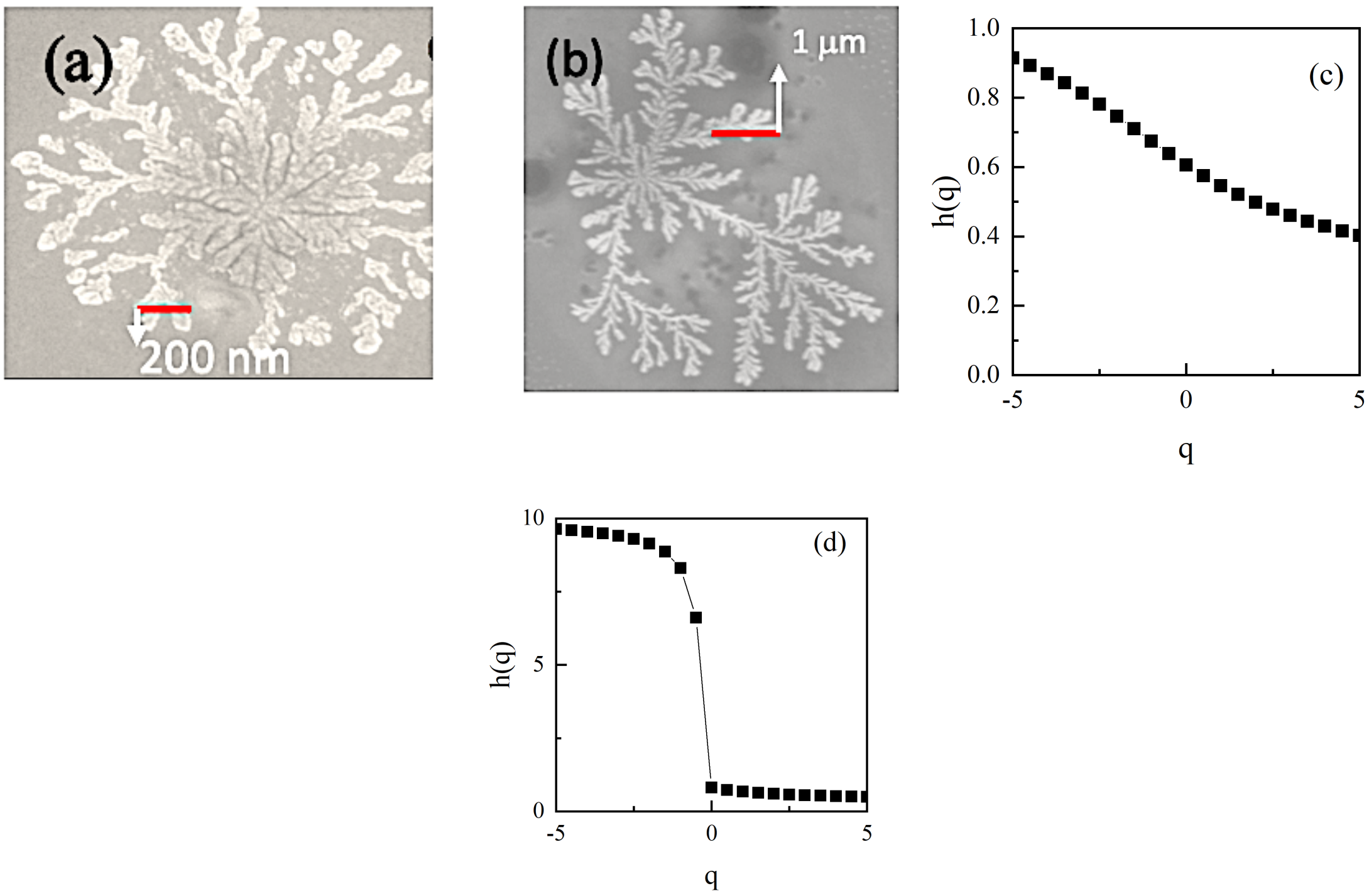

Figure 3: FE-SEM images of NbN films of 300 nm thickness deposited on Si substrate at a substrate temperature of 250°C, (a) sample 5 at a magnification of 91.65 KX, and (b) sample 6 at a magnification of 29.82 KX. The variation in the corresponding Hurst exponents h(q) as a function of the moments, q are shown in (c) and (d) respectively.

Table 2: Hurst exponent and the monofractal Hausdorff dimension for the different samples

| | Hurst Exponent (H) | Correlation | Hausdorff dimension |
|---|---|---|---|
| Sample 1 | 0.76 | Persistent (long-range correlation) | 1.75 |
| Sample 2 | 0.62 | Persistent (long-range correlation) | 1.65 |
| Sample 3 | 0.26 | Anti-persistent | 1.79 |
| Sample 4 | 0.50 | No correlation (Random) | 1.29 |
| Sample 5 | 0.60 | Persistent (long-range correlation) | 1.74 |
| Sample 6 | 0.23 | Anti-Persistent | 1.64 |

In 2D MFDFA analysis, at the lower end, $\Delta\alpha = 0.35$ suggests a surface or image that is nearly monofractal or weakly multifractal, characterized by relatively uniform structures with minor variations in local scaling behaviour. At the higher end, $\Delta\alpha > 1$ reflects extremely strong multifractality, indicating a highly heterogeneous surface with a broad spectrum of local roughness or intensity variations. This broad range implies that the system exhibits diverse structural complexity, likely influenced by factors such as temperature, deposition conditions, or material type. Surfaces with low Δα tend to be smoother and more homogeneous, whereas those with high Δα are more irregular, hierarchical, and scale-rich [35].

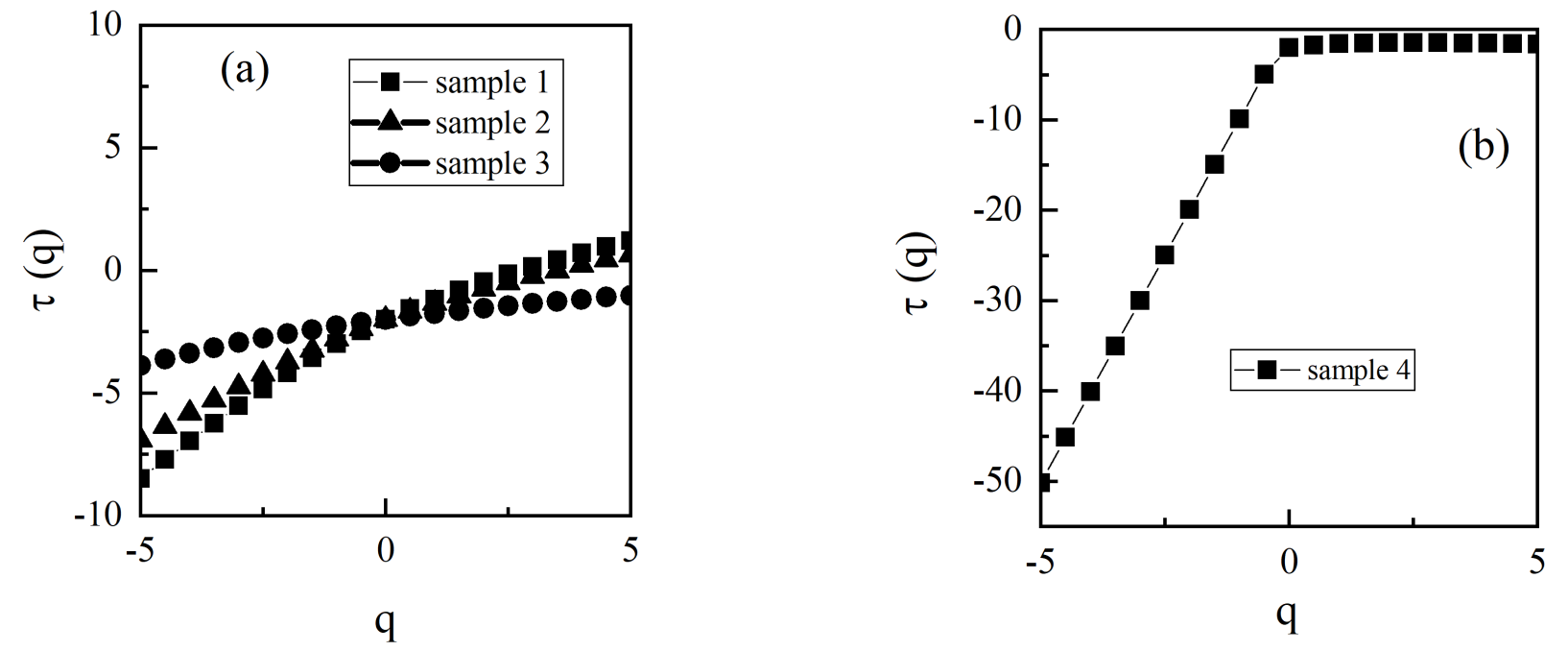

Figure 4: The mass exponent of (a) samples 1, 2 and 3 and (b) sample 4 plotted as a function of the moment, q.

The multifractal singularity spectrum *f(α)* versus *α* is presented in respectively in Figs. 6(a), (b), (c) and (d) for samples 1,2,3 and 4 and figs 7(a) and (b) for samples 5 and 6 respectively. The variation of *f(α)* vs *α* is expected to be an asymmetric bell-shaped curve whose width $\Delta\alpha = \alpha_{max} - \alpha_{min}$ is sample dependent. The Δα values are positive and in the range of 0.35 to 1.05 for samples 1, 2, 3 and 5. The Δα values for samples 4 and 6 could not be determined due to poor image resolution. However, it is evident that the maximum *f(α)* value is ~2 which is the same as that for the other samples. Talu et al. have reported values in the range of 0.2023 to 1.007 for TiN thin films [25], 2.99 or greater for ITO thin films[39], 0.0007 to 0.167 for CuTsPc films deposited on quartz and glass substrates [40] and Das et al. have reported values of the order of 2 for Ti Thin films [41].

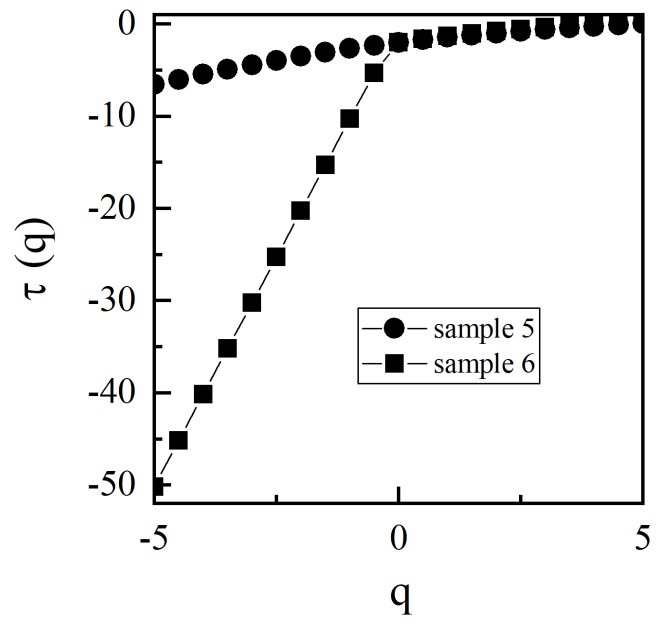

Figure 5: The mass exponent of samples 5 and 6 plotted as a function of the moment, q.

It is evident from the SEM images presented, that a wide variety of fractal structures are formed. An aspect that was not examined closely, in our previous works, was the presence of multifractality in these structures.

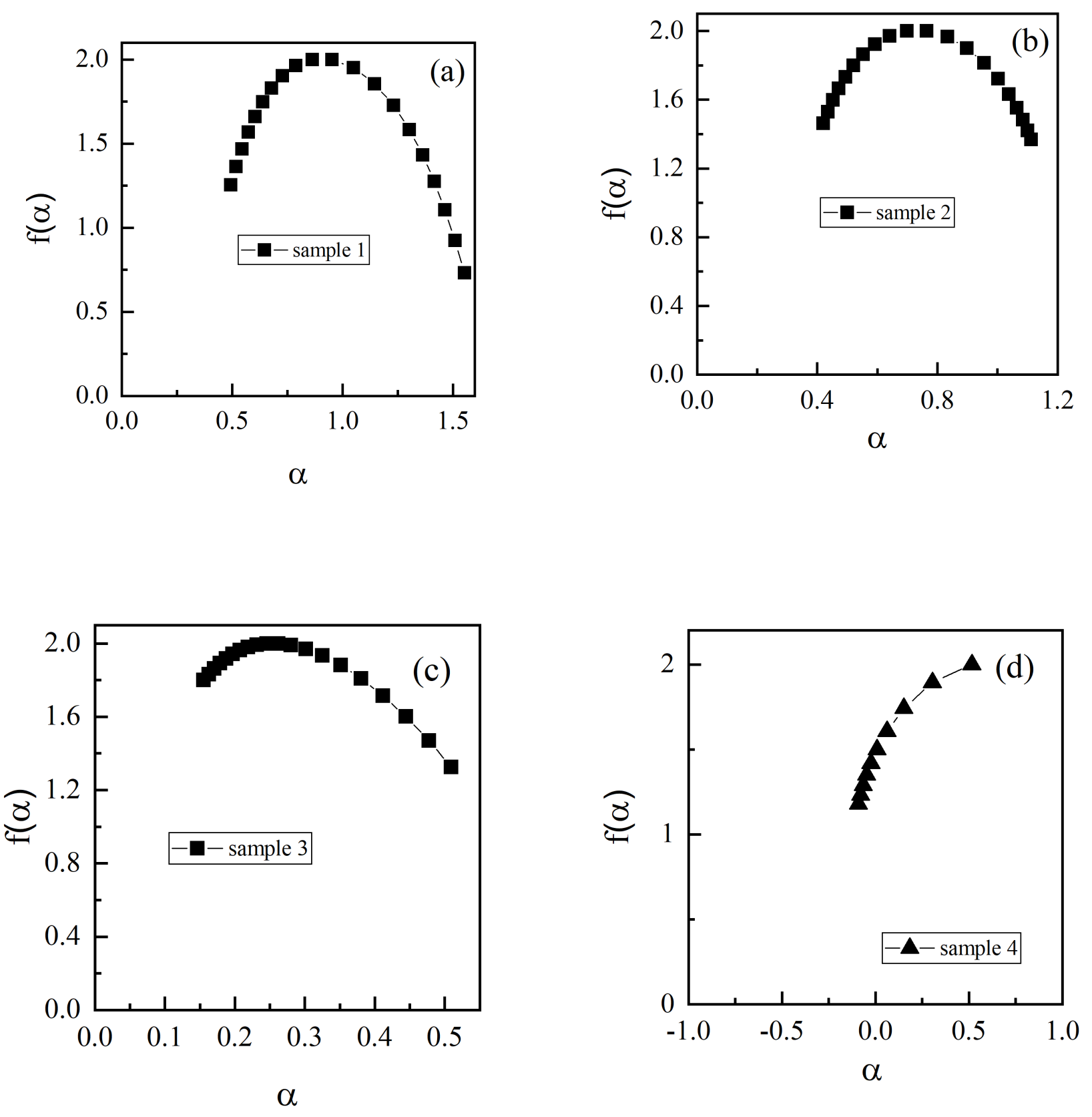

Figure 6: The multifractal spectrum of f(α) against α, where α represents the strength of singularities, and f(α) is the fractal dimension of points with those singularities for (a) sample 1, (b) sample 2, (c) sample 3 and (d) sample 4.

Possible physical interpretations of the results can be drawn by examining how multifractal parameters vary with temperature. The variation of $\Delta\alpha$ with temperature may reflect changes in surface diffusion, where higher temperatures enhance diffusion and reduce heterogeneity in growth, potentially narrowing the spectrum (smaller $\Delta\alpha$). Conversely, thermally activated instabilities could broaden the spectrum (larger $\Delta\alpha$) by generating more complex, scale-rich structures. The Hurst exponent (H) as a function of temperature can indicate smoothing effects, as elevated temperatures may promote surface relaxation, leading to reduced H values and a tendency toward anti-persistent behaviour.

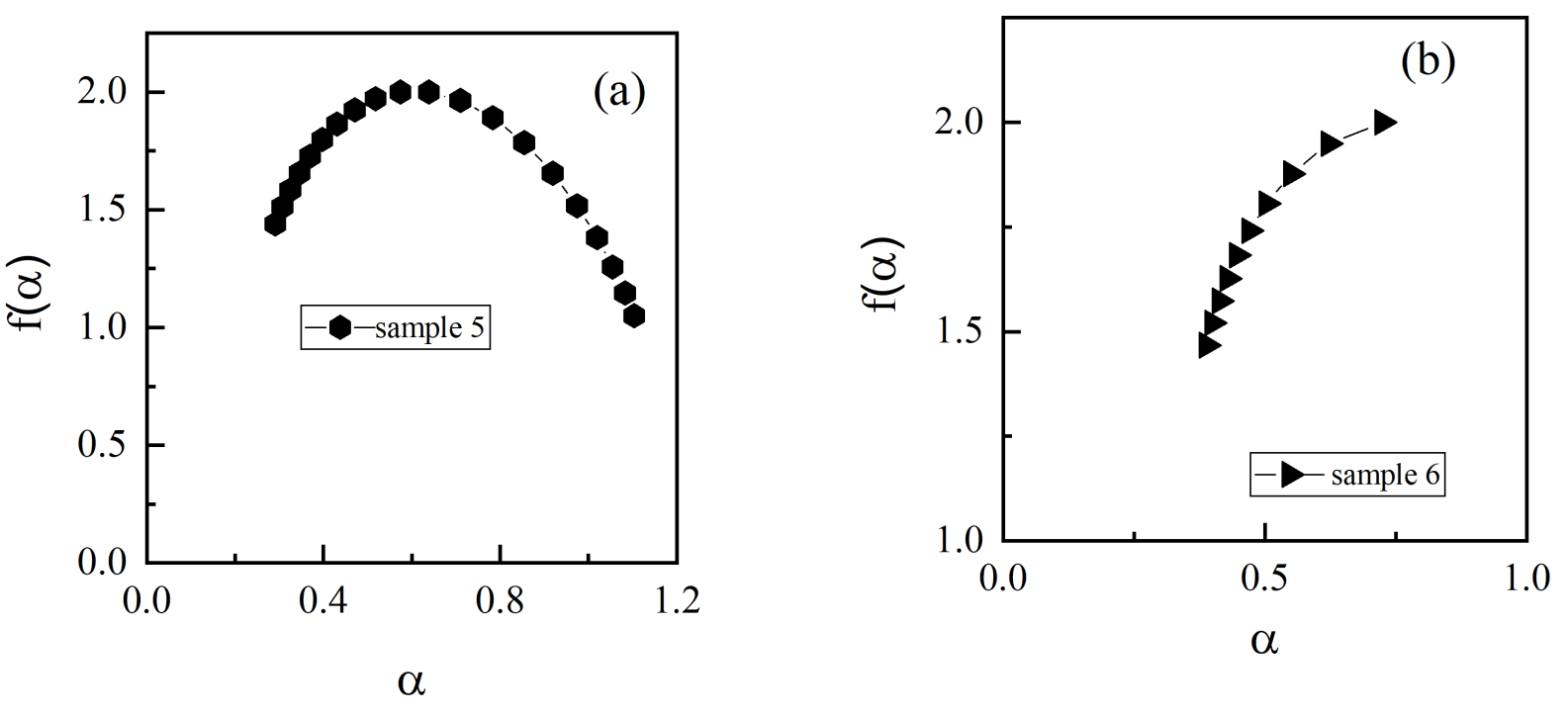

Figure 7: The multifractal spectrum of f(α) against α, where α represents the strength of singularities, and f(α) is the fractal dimension of points with those singularities for (a) sample 5, (b) sample 6.

This shift may suggest a transition from kinetic roughening to thermally dominated smoothing. Similarly, the fractal dimension ($D_f$) provides insights into surface morphology, where an increase in $D_f$ with rising temperature could signify greater surface complexity due to dynamic rearrangements, while a decrease in $D_f$ may correspond to smoother, more uniform surfaces [1-5]. Crystalline films develop through nucleation and grain growth, producing grain boundaries, anisotropic stresses, and defects that generate long-range correlations and hierarchical structures, thereby enhancing multifractality. Nucleation theory explains how variations in grain size, rate, and orientation contribute to heterogeneous morphology [23, 24]. In contrast, amorphous films grow isotropically without grain boundaries, resulting in smoother, more homogeneous surfaces where random deposition, diffusion, and shadowing dominate, typically leading to monofractal or weakly multifractal behaviour.

In the present work, each parameter reported (e.g., $h(q)$, $H$, $Dq$, $D_f$, and $\Delta\alpha$) was obtained from representative regions, and consistent scaling behaviour was verified across the analyzed images. Consistency has been ensured by applying a uniform pre-processing pipeline (a global threshold with fixed parameters across all samples) and verifying that moderate variations in thresholding or filtering do not qualitatively alter the observed multifractal scaling. The reproducibility of these trends among independently prepared samples was found to be satisfactory within experimental scatter and also benchmarked with literature [42-48].

The Hausdorff dimension quantifies surface complexity or roughness across scales, ranging from 1–2 for 1D profiles and 2–3 for 2D surfaces. Higher value indicates more irregular, fractal surfaces, affecting physical properties: In sputtering, the Hausdorff dimension can be tuned via ion energy, deposition angle, and substrate temperature, allowing surfaces to be optimized [15, 23, 25].

It is generally accepted in literature that Hausdorff dimension values of the order of 1.7 are attributable to diffusion limited aggregation[49-51]. Early work by Stanley and Meakin [49] as well as Witten and Sander [50] showed that multifractals have a dimension of ~1.7 and also attributed the presence of multifractals to diffusion-limited aggregation (DLA). More recently, Tenti et al. have demonstrated that diffusion limited aggregation leads to fractal dimension of the order of 1.7 [51]
In the specific case of magnetron sputtered TiN films, Talu etal [25] have reported that the fractal dimension increased from 2.36 at 473 K to 2.66 at 673 K and then decreased to 2.33 at 773 K. Korpi et al. [26] and Perry [27] have reported values of Hausdorff dimension of the order of 1.7-2 for TiN thin films.
It is, thus, evident that Hausdorff dimension of the samples in the present case falls in the regime that can be attributed to DLA. The differences in values are attributed to differences in processing conditions, substrate type and other parameters. In DLA diffusion of the species toward the surface (or heat away from it) can be the rate limiting process. The aggregates display multibranched forms which are also observed in objects such as dust balls, agglomerated soot, and dendrites. The origin of these complex shapes is attributed to the diffusion-limited process induced instabilities. It is pointed out that Meakin [1], Stanley and Meakin [49] , Witten and Sander [50], and Schwarzer et al. [52] have suggested that branching as well as directionality in these structures is a consequence of the DLA process. In the present work, it appears that the elastic strain at the film-substrate interface also contributes to the directionality. The proposed models were, therefore, able to reproduce the experimentally observed multibranched structures. Furthermore, the correlations that arose between particle positions in the model are reminiscent of a scale-invariant object. It is, pertinent to note that the origin of elastic strain in the case of films deposited on amorphous substrates is the difference in thermal expansion coefficients. In the case of films deposited on single crystal substrates, however, the elastic strain is due to the difference in lattice constants. These two parameters create the necessary conditions at the film-substrate interface leading to morphological instabilities [28]. Therefore, it is proposed that the competition between substrate temperature and the strain at the film-substrate interface results in morphological instabilities and diffusion limited aggregation.

## 5. Summary

Multifractal structures in RF magnetron sputtered thin films of TiN and NbN deposited on single crystal Si and amorphous glass substrates are presented. The multifractals comprise dendrites and snow-flakes with self-similarity over several length scales ranging from tens of microns to tens of nanometers. Furthermore, multifractality is clearly dependent on initial conditions of growth such as substrate temperature, pressure and the nature of the film-substrate interface. The 2D Multifractal Detrended Fluctuation Analysis (2D MFDFA) method is used to quantify parameters such as Hausdorff coefficient, Hurst exponent, fractal dimensions and width of the multifractal singularity spectrum. It is proposed that the origin of these structures is a combination of morphological instabilities and diffusion-limited aggregation.

**Acknowledgements**
The authors acknowledge the characterization and computing facilities provided by the University of Hyderabad. No external funding was received for this research work. The authors thank Ms. Sharon Jackson, a well-known violinist from Los Angeles, who helped us improve our understanding of the role of fractals in music transmission.